\documentclass{iaus}
\usepackage{graphicx}

\title[Physical State of the Deep Interior of the CoRoT-7b Exoplanet] 
{Physical State of the Deep Interior of the CoRoT-7b Exoplanet}

\author[F.W. Wagner, F. Sohl, T. R{\"u}ckriemen, H. Rauer]   
{Frank W. Wagner$^1$, Frank Sohl$^1$, Tina R{\"u}ckriemen$^{1}$, Heike Rauer$^{1,2}$}

\affiliation{$^1$Institute of Planetary Research, German Aerospace Center (DLR), Berlin, Germany \\[\affilskip]
$^2$Center of Astronomy and Astrophysics, Berlin Institute of Technology, Berlin, Germany}

\pubyear{2010}
\volume{276}  
\pagerange{1--4}
\addtocounter{figure}{0}
\jname{The Astrophysics of Planetary Systems: \\ Formation, Structure, and Dynamical Evolution}
\editors{A. Sozzetti, M.G. Lattanzi, and A.P. Boss, eds.}
\begin{document}
\maketitle
\begin{abstract}
The present study takes the CoRoT-7b exoplanet as an analogue for massive terrestrial planets to investigate conditions, under which intrinsic magnetic fields could be sustained in liquid cores.
We examine the effect of depth-dependent transport parameters (e.g., activation volume of mantle rock) on a planet's thermal structure and the related heat flux across the core mantle boundary.
For terrestrial planets more massive than the Earth, our calculations suggest that a substantial part of the lowermost mantle is in a sluggish convective regime, primarily due to pressure effects on viscosity.
Hence, we find substantially higher core temperatures than previously reported from parameterized convection models.
We also discuss the effect of melting point depression in the presence of impurities (e.g., sulfur) in iron-rich cores and compare corresponding melting relations to the calculated thermal structure.
Since impurity effects become less important at the elevated pressure and temperature conditions prevalent in the deep interior of CoRoT-7b, iron-rich cores are likely solid, implying that a self-sustained magnetic field would be absent.
\end{abstract}

\firstsection 
\section{Introduction}

\begin{figure}[ht!]
\begin{minipage}{0.60\textwidth}
\begin{center}
  \includegraphics[width=\textwidth]{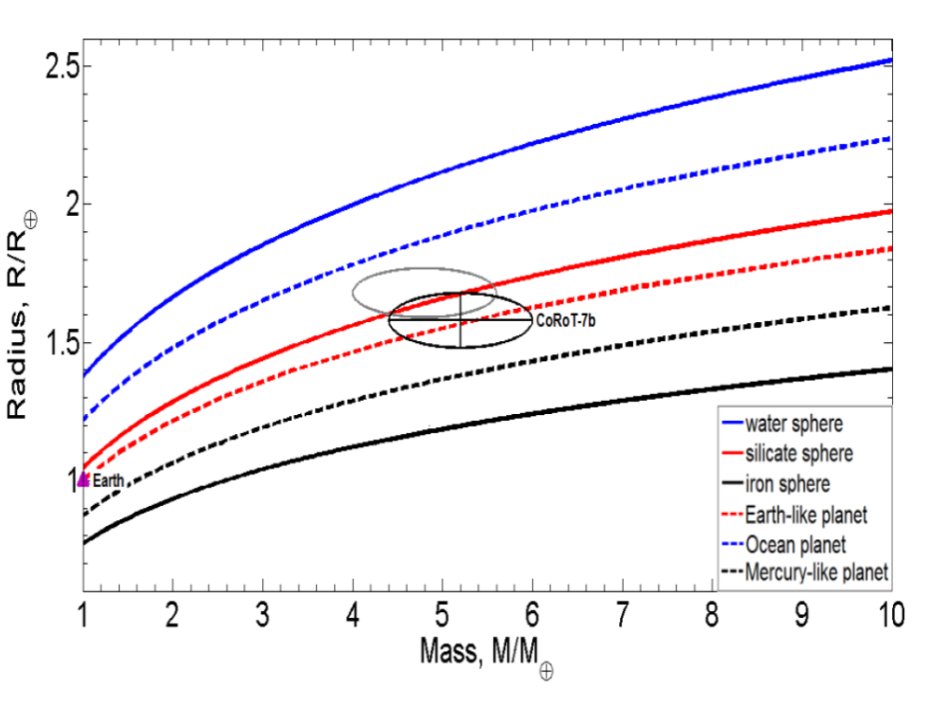}\\
  \label{mr10}
\end{center}
\end{minipage}
\begin{minipage}{0.40\textwidth}
\begin{center}
  \caption{\small Mass-radius relations for differentiated exoplanets (dashed lines) and homogeneous, self-compressible spheres of water, silicate, and iron (solid lines) ranging from $1$ to $10$~M$_\oplus$. The relative position of the Earth is indicated by the triangle. Whereas the gray ellipse denotes the previously reported range for total mass $M_{p}$ (\cite{7}) and planetary radius $R_{p}$ (\cite{1}), the black ellipse represents the recently revised range for $R_{p}=(1.58\pm0.10)$~R$_{\oplus}$ and $M_{p}=(5.2\pm0.8)$~M$_{\oplus}$ (\cite{27}).}
\end{center}
\end{minipage}
\end{figure}

CoRoT-7b is probably the most prominent discovery of the CoRoT space mission as it is the first extrasolar planet below ten Earth masses with firm observational constraints on planetary mass {\it and} radius.
Although the planet's exact mass is still subject to debate (e.g, \cite{34}, \cite{35}), a total mass of about five times that of the Earth seems to be most favored (\cite{27}, \cite{7}).
Combined with the measured radius of ($1.58\pm0.10$)~M$_\oplus$ (\cite{27}), this suggests an average compressed density of ($7.2\pm1.8$)~Mg~m$^{-3}$ for CoRoT-7b, being consistent with a terrestrial bulk composition (Fig.~\ref{mr10}).
In the following, we take CoRoT-7b as a type example for a terrestrial exoplanet to investigate the physical state and thermal structure of planetary interiors.
The purpose of this study is to understand the necessary conditions, under which an intrinsic magnetic field could be sustained in liquid cores of massive terrestrial planets.

\firstsection
\section{Method}

We use a one-dimensional, four-layer structural model that is similar to the model approach disclosed on the occasion of the XXVIIth IAU General Assembly held in Rio de Janeiro, Brazil (\cite{18}).
The interior structure of CoRoT-7b is obtained by solving the mass and energy balance equations in conjunction with an equation of state (EoS) for the internal density distribution.
The Birch-Murnaghan EoS (\cite{16}) and the Rydberg-Vinet EoS (\cite{15}) are often criticized because of inconsistencies in respect to the physics of the thermodynamic limit (e.g., \cite{17}).
Hence, we improve previous models (e.g., \cite{3}) by implementing a generalized Rydberg EoS (\cite{19}), facilitating extrapolation to exceptionally high pressures (\cite{14}).

We adopt a mixing length formulation (\cite{4}) to self-consistently calculate the thermal state of planetary mantles.
This is an improvement to our previously proposed model, in which we have used this approach only for the upper mantle to simulate a lithosphere within the uppermost part (\cite{18}).
The basic idea behind the mixing length concept is that internally generated heat is primarily transferred by vertical motion of fluid parcels which will entirely loose their individuality after migrating across size-dependent characteristic length scales.
\cite{22} demonstrated the feasibility of that concept for modeling of heat transfer within terrestrial planet interiors.
The viscosity determining the temperature profile is modeled in the framework of a temperature- and pressure-dependent Arrhenius viscosity law, considering diffusion and dislocation creep as the most important creep mechanisms in silicate aggregates (\cite{31}).
The pressure-induced reduction of activation volume with depth is approximated as a vacancy in the material (\cite{30}).
Solidification temperatures for the silicate-dominated mantle are obtained from ab initio molecular dynamics simulations (\cite{23}).
To construct melting relations under core conditions for pure iron and iron-rich alloys containing impurities, we use a parameterization based on the theory of tricritical phenomena (\cite{24}) up to $800$~GPa and extrapolate according to the well-known Lindemann law (\cite{25}).
\firstsection
\section{Results and Discussion}

\begin{figure}[ht!]
\begin{center}
  \includegraphics[width=\textwidth]{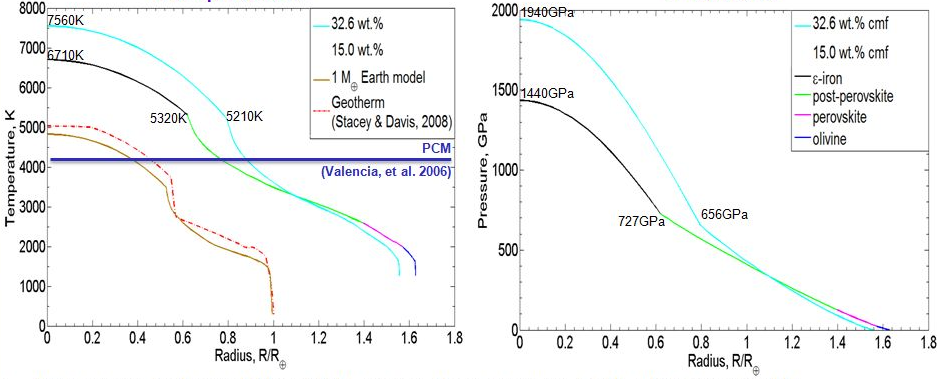}\\
  \caption{\small Interior structure of CoRoT-7b corresponding to different iron core mass fractions: (Left) Radial distribution of temperature. An Earth-sized model and a reference geotherm (\cite{32}) are shown for comparison. (Right) Radial distribution of hydrostatic pressure.}
  \label{b33}
\end{center}
\end{figure}

First, we present modeling results as obtained by using CoRoT-7b as case study and calculating self-consistently the thermal structure within planetary mantles and cores.
Figure~\ref{b33} (left-hand side) shows the radial temperature profile of two plausible CoRoT-7b models of $5$~M$_{\oplus}$ that differ in the iron core mass fraction assumed, representing Earth-like (i.e., $32.6$~wt.-\%) and iron-depleted (i.e., $15.0$~wt.-\%) bulk compositions.
From a fixed surface value of $1270$~K (\cite{1}), temperature increases rapidly across a thin thermal lithosphere, where heat is transferred by conduction and convection is absent.
A mostly adiabatic temperature rise is observed across the convecting mantle and underlying core, resulting for the Earth-like and iron-depleted case, respectively, in temperatures of $5210$~K and $5320$~K at the core-mantle boundary (CMB) and up to $7560$~K and $6710$~K in the planet's center.
Compared to parameterized convection models (e.g., \cite{10}), we find generally higher CMB temperatures.
This is caused by a steeper temperature gradient prevalent in the deep interior that is mainly attributed to the pressure-induced increase of viscosity at elevated lower mantle pressures and temperatures.
To illustrate environmental conditions in the deep interiors of massive exoplanets, the corresponding pressure profiles for the two CoRoT-7b cases are shown in Fig.~\ref{b33} right-sided.
Whereas CMB pressures of CoRoT-7b are twice as large as the central pressure of the Earth, the central pressure is as much as five times the pressure at the center of the Earth, depending on the planet's iron core mass fraction.

\begin{figure}[ht!]
\begin{center}
  \includegraphics[width=0.70\textwidth]{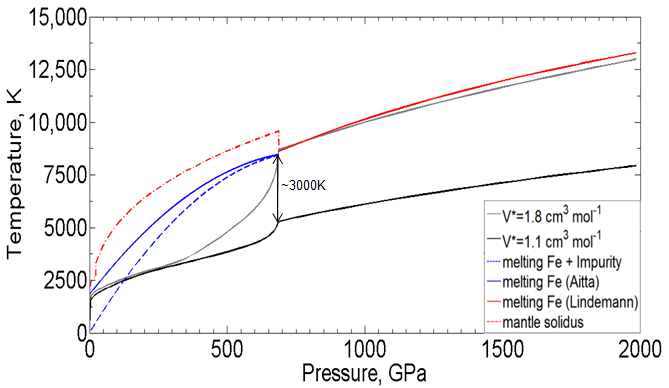}
  \caption{\small Influence of activation volume $V^{*}$ and core impurities (e.g., sulfur) on the physical state of matter within a $5$~M$_{\oplus}$ exoplanet.~ ~ ~ ~ ~ ~ ~ ~ ~ ~ ~ ~ ~ ~ ~ ~ ~ ~ ~ ~ ~ ~ ~ ~ ~ ~ ~ ~ ~ ~ ~ ~ ~ ~ ~ ~ ~ ~ ~ ~ ~ ~ ~ ~ ~ ~ }
  \label{b44}
\end{center}
\end{figure}

Next, we investigate conditions for the possible existence of liquid cores, which are prerequisite for magnetic field generation.
In Figure~\ref{b44}, pressure-temperature ($P$-$T$) profiles for two Earth-like CoRoT-7b models are shown together with the corresponding silicate and iron melting relations.
It is seen that iron melting temperatures are generally lower than solidus temperatures of MgSiO$_{3}$ (post-)perovskite under CMB conditions.
The temperature profile shown in black corresponds to the Earth-like model case illustrated in Fig.~\ref{b33} and yields an activation volume of $1.1$~cm$^{3}$~mol$^{-1}$ at the CMB.
Lower mantle and core materials are expected to be solid because of extremely high compression at relatively low temperature.
A temperature increase of about $3000$~K at the CMB would be required to facilitate melting of the outer core.
The temperature profile shown in gray corresponds to the latter case, for which a larger activation volume of $1.8$~cm$^{3}$~mol$^{-1}$ at the CMB would be required.
Hence, the lower mantle becomes stiffer with increasing activation volume and, as a consequence, local temperature will rise until intersection with the iron melting curve.
Also notable in that respect is that the melting intervall in the presence of impurities (e.g., siderophile elements like sulfur) would vanish at extremely high pressures (blue curves), according to (\cite{24}).
Due to large pressures within the deep interior of a $5$~M$_{\oplus}$ exoplanet such as CoRoT-7b, the addition of impurities will not modify much the $P$-$T$ relation compared to that of a pure iron core.
This suggests that impurities should only have a minor effect on core melting within massive terrestrial exoplanets (\cite{14}).

\firstsection
\section{Conclusions}
It is concluded that the physical state of the deep interior of massive terrestrial exoplanets is strongly dependent on mantle rheology.
Impurities have only a minor effect on core melting, which can be explained conceptually by the large compression of matter under core conditions.
Due to the large effect of pressure on melting, a pure iron core is expected to be solid and, therefore, a self-generated magnetic field should be absent on massive terrestrial exoplanets like CoRoT-7b.
Nevertheless, liquid cores cannot completely be ruled out, but in that case substantially larger activation volumes ($>1.8$~cm$^{3}$~mol$^{-1}$) compared to values predicted by the vacancy approach would be required to initiate core melting.
Furthermore, a pressure-induced sluggish convection is prevalent in the lowermost mantle and should influence mantle convection pattern as well as the thermal evolution of massive exoplanets.

\begin{acknowledgment}
\noindent
This research is supported by the Helmholtz Alliance "Planetary Evolution and Life".
\end{acknowledgment}
\firstsection

\end{document}